\documentclass[12pt,a4paper]{elsarticle}
\usepackage{amsfonts,physics}
\usepackage[utf8]{inputenc}
\usepackage{hyperref}
\hypersetup{colorlinks=true,urlcolor=red, citecolor=blue,linkcolor=blue}
\allowdisplaybreaks
\usepackage{amsmath}
\usepackage{amsfonts}
\usepackage{amssymb}
\usepackage{graphicx,epstopdf}
\usepackage{float}
\begin{document}
	\title{Nonclassical properties of a deformed atom-cavity field state}
	\author{Naveen Kumar}
	\ead{naveen74418@gmail.com}
	\author{Deepak}
	\ead{deepak20dalal19@gmail.com}
	\author{Arpita Chatterjee\corref{cor1}}
	\ead{arpita.sps@gmail.com}
	\cortext[cor1]{Corresponding author}
	\date{\today}
	\address{Department of Mathematics, J. C. Bose University of Science and Technology,\\ YMCA, Faridabad 121006, India}
	\begin{abstract}
		We analyze here a nonclassical state produced by an atom-cavity field interaction. The two-level atom is passed through the single-mode electromagnetic cavity field. By deforming the field operators and introducing nonlinearity to the classic Jaynes–Cummings model, we explore the system in respect of a nonlinear Hamiltonian. Assuming that the atom is in an excited state and the field is in a coherent state initially, the analytic expression for the state vector of the entire system state vector is obtained. With the help of the derived state vector, we calculate the photon number distribution, Mandel's $Q_M$ parameter, Wigner function, anti-bunching, squeezing properties and $Q$ function etc.
	\end{abstract}
	\maketitle
	\section{Introduction}
	The conventional Jaynes–Cummings model (JCM) \cite{jcm} describes the atom-cavity interaction as a two-level atom is colliding with a single mode of the electromagnetic field in the matter–radiation coupling. This model is proposed as a basic design to investigate the semiclassical behaviour of quantum radiation field \cite{rad}. The generalized Jaynes-Cummings model, illustrating the nonlinear interaction of a two or multi-level atom with a cavity field, results a deformed JCM
structure. In quantum optics and quantum information processing \cite{it}, the nonclassical light field is of major interest for a number of reasons. In an all-optical quantum information processing device \cite{d}, the single-photon Fock state, a nonclassical state, is an essential resource. Controlling the emission of a single radiator, such as a molecule or a quanta \cite{brunal}, can be used to create these states. Fock state can also be prepared using cavity QED experiments in which atoms interact one at a time with a high $Q$ resonator. A $\pi$ quantum Rabi pulse in a microwave cavity \cite{hagley} or an adiabatic passage sequence in an optical cavity\cite{henrich} can produce a one-photon Fock state in this way. The study of these states yields a fundamental understanding of quantum fluctuations and a new method of quantum communication or imaging that surpasses the standard quantum noise limit. Nonclassical states have a wide range of real-world applications. For example,
squeezed states are used to reduce the noise level in
one of the phase-space quadratures below the quantum
limit \cite{ql}, entangled states are employed to realize a
quantum computer and to transfer quantum information \cite{qi}. Here under the rotating-wave approximation (RWA), we investigate the dynamics of two-photon correlations generated by the interaction of a semiclassical two-level atom with a single-mode cavity field.

The following is a breakdown of the paper's structure. We begin by describing the interaction between a two-level atom and a single-mode field and
solve the time-dependent Schr\"{o}dinger equation \cite{ql}, and derive the definite form of the atom-cavity state vector. To find the field's nonclassical properties, we first trace out the atom part from the atom-field system. We then check the nonclassicality criteria for distribution of photon numbers that means the probability of detecting $n$ photons in the state $\hat{\rho}$. It is a smooth function of $n$ for classical states like coherent and thermal states and an oscillating function of $n$ for nonclassical states like squeezed states, as described in Schleich and Wheeler's \cite{pn} seminal work on interference in phase space. We further discuss the nonclassicality of the cavity field described by Mandel’s $Q_M$ parameter, Wigner function, anti-bunching, squeezing parameter, and $Q$ function.

	\section{State of interest}
	In a typical Jaynes-Cummings model, a two-level atom interacts with a single-mode quantized electromagnetic field oscillating with frequency $\Omega $. We consider $\sigma_{ij}=\ket{i}\bra{j}, \,(i,j=1,2)\,$ with $\ket{2}$ ($\ket{1}$) as the excited (ground) state of the atom. With these notations, the resonant Jaynes-Cummings Hamiltonian can be written as (with units such that $\hbar=1$) \cite{siva, deformed}
	\begin{align}
		\label{ijcm}
		H= \Omega N+ \omega \sigma_3+ g(a \sigma_{21}+a^\dag \sigma_{12})
	\end{align}
where $N=a^\dag a$ is the number operator for the electromagnetic field mode, $\sigma_{3}=\ket{2}\bra{2}-\ket{1}\bra{1}$ is the atomic inversion operator, $\omega=\omega_2-\omega_1$ is the atomic transition frequency, and the coupling constant $g$ is effectively the product of the dipole moment and the mode
function of the cavity mode at the position of the atom. The first two terms of \eqref{ijcm} are the free Hamiltonians of
a single-mode electromagnetic field and a two-level atom. When the atom is interacting with the cavity field, it
may drop (jump) from the excited (ground) to ground (excited) state by emitting (absorbing) a photon. The Hamiltonian \eqref{ijcm} describes one of the few exactly solvable models in quantum optics. The deformation can be added to the atom-field system in two different ways. First, the field properties can be modified by an intensity-dependent coupling between the atom and the radiation field, i.e. the coupling is no longer linear in the field variables. Secondly, a system is considered in which the electromagnetic field mode is excited in a Kerr medium. This system may be realized by the experiment with a Rydberg
atom in a nonlinear Kerr-like cavity. We consider here the first kind of nonlinearity only. The Hamiltonian describing the dynamics of this quantum system in the rotating wave approximation (RWA) is obtained from \eqref{ijcm} by deforming the bosonic operators as
	\begin{equation}
		\label{mh}
		H=\Omega {\hat{A}}^{\dagger}\hat{A}+(\omega_1\sigma_{11}+\omega_2\sigma_{22})+g(\hat{A}\sigma_{21}+{\hat{A}}^\dagger \sigma_{12}).
	\end{equation}
	The $f$-deformed annihilation ($\hat{A}$) and creation (${\hat{A}}^{\dagger}$) operators are constructed from the
usual bosonic operators $\hat{a}$, ${\hat{a}}^{\dagger}$ and the number operator
$\hat{n} = a^\dag a$ such as ${\hat{A}}^{\dagger}=f(\hat{n})a^\dag$ and $\hat{A}=\hat{a} f(\hat{n})$, where $f(\hat{n})$ is an arbitrary real function of the number operator $\hat{n}$ \cite{deformed1,deformed2}. One way of realizing nonlinear Jaynes-Cummings Hamiltonian is to consider atomic systems
with vibrational sidebands \cite{filo}. The spatial structure of the cavity field determines the form for the
function $f(a^\dagger a)$. By proper choice of cavity mode structure, it is possible to design
arbitrary atom–field couplings.
Another way to arrive at atom-field coupling, which has polynomial dependence on the photon number, is to use many lasers with different phases and Rabi
frequencies to interact with a trapped ion with vibrational sidebands \cite{siva}. The number
of lasers required is same as the order of the polynomial. Once the specific form of the function $f(n)$, considered to be a polynomial, is fixed, the required
phases and the Rabi frequencies of the external lasers can be determined. With the technological advancement in trapping and availability of lasers, it is possible to engineer the atom–environment coupling
by varying laser frequencies and their intensities.
%  \cite{deformed2}. {\color{blue} On the basis of of parameters,  coupling constant $(g)$, energies $ (w_1,w_2)$ we can differentiate the coupling regime in three ways namely strong coupling regime ($g/w_i\lesssim{10}^{-2} $), Ultra-strong coupling regime ($ g/w_i\gtrsim 10^{-1} $), and deep strong coupling ($ g/w_i\approx1 $) with $ i=1,2$ \cite{am12,am14}.}
	Some $f$-deformed bosonic commutations relations are
	\begin{align*}
		[\hat{A},{\hat{A}}^{\dagger}]&=(\hat{n}+1)f^2(\hat{n}+1)-\hat{n}f^2(\hat{n})=g(\hat{n}),\,\textrm{say}\\ [{\hat{A}}^{\dagger}\hat{A},\hat{A}]&=-g(\hat{n})\hat{A},~~~~~~~~~~~
		[{\hat{A}}^{\dagger}\hat{A},{\hat{A}}^{\dagger}]={\hat{A}}^{\dagger}g(\hat{n})\\
		[{\hat{A}}^{\dagger}\hat{A},{\hat{A}}^{\dagger}g(\hat{n})]&={\hat{A}}^{\dagger}g^2(\hat{n}),~~~~~[{\hat{A}}^{\dagger}\hat{A},{\hat{A}}^{\dagger}g^p(\hat{n})]={\hat{A}}^{\dagger}g^{p+1}(\hat{n}) \\
		[{\hat{A}}^{\dagger}\hat{A},-g(\hat{n})\hat{A}]&=g^2(\hat{n})\hat{A},~[{\hat{A}}^{\dagger}\hat{A},(-1)^p g^p(\hat{n})\hat{A}]=(-1)^{p+1}g^{p+1}(\hat{n}) \hat{A}
	\end{align*}
	The Hamiltonian in the interaction picture is
	\begin{eqnarray*}
		H_r & = & e^{iH_{r_0}t}H_{r_1} e^{-iH_{r_0}t}
	\end{eqnarray*}
	where
	\begin{align*}
		H_{r_0}&=\Omega {\hat{A}}^{\dagger}\hat{A}+w_1\sigma_{11}+w_2\sigma_{22}\\
		H_{r_1}&=g(\hat{A}\sigma_{21}+{\hat{A}}^\dagger \sigma_{12})
	\end{align*}
	Applying the Baker-Campbell-Hausdorff formula, $H_r$ can be calculated as
	\begin{eqnarray}\nonumber
		\label{ih}
		H_r & = & H_{r_1}+(it)[H_{r_0},H_{r_1}]+\frac{(it)^2}{2!}[H_{r_0},[H_{r_0},H_{r_1}]]
		+\frac{(it)^3}{3!}[H_{r_0},[H_{r_0},[H_{r_0},H_{r_1}]]]+\ldots
		\\
& = &\sum_{r=0}^\infty\frac{(it)^r}{r!}[H_0,[H_0,[H_0,...H_1]]]_{r\text{ times}}
\end{eqnarray}
and thus the effective Hamiltonian as
\begin{eqnarray}
		H_{\mathrm{eff}}&=& g [e^{-ith(\hat{n})}A\sigma_{21}+A^\dagger e^{ith(\hat{n})}\sigma_{12}]
	\end{eqnarray}
	where $h(\hat{n})=\Omega f^2(\hat{n}) +(\omega_2-\omega_1)$.\\
	Solving the Schr\"{o}dinger equation of motion \cite{ql}
	\begin{eqnarray}
		i\frac{\partial | \psi \rangle }{\partial t}= H |\psi \rangle,
	\end{eqnarray}
for any arbitrary state $|\psi\rangle = \sum {C_{1, n+1}(t) |1, n+1\rangle +C_{2, n}(t) |2, n\rangle}$, $\ket{2}$ and $\ket{1}$ are the excited and ground states of atom, respectively, the equations of motion in terms of the probability amplitudes of the state vector $\ket{\psi}$ are
\begin{eqnarray*}
	\label{dre}
	\dot{C}_{2,n+1}(t) & = & e^{ih(\hat{n})t}\sqrt{n+1}\,C_{1,n}(t),\\\\
	\dot{C}_{1,n}(t) & = & e^{-ih(\hat{n})t}\sqrt{n+1}\,C_{2,n+1}(t).	
\end{eqnarray*}
Assuming that the atom enters the cavity in its excited state $\ket{2}$ and the
radiation field is initially in a coherent state, that means $C_{1,n}(0) = 0$ and $C_{2,n+1}(0) = C_n(0) = e^{-\frac{|\beta|^2}{2}}\frac{\beta^n}{\sqrt{n!}}$,
the above equations can be solved as

\begin{align}
	\label{x1}
	C_{2,n+1}(t)&=\frac{e^{-\frac{|\beta|^2}{2}}\beta^n}{{\sqrt{n!}}(m_2-m_1)}\left\{m_2e^{m_1t}-m_1e^{m_2t}\right\}\\
	\label{x2}
	C_{1s,n}(t)&=\frac{e^{-ih(n)t}m_1m_2e^{-\frac{|\beta|^2}{2}}\beta^n}{\sqrt{n!}\sqrt{n+1}gf(n+1)(m_2-m_1)}\left\{e^{m_1t}-e^{m_2t}\right\}
\end{align}
where
\begin{align*}
	2m_{1, 2}=ih(n)\pm i\sqrt{4g^2(n+1) f^2(n+1)+h^2(n)}
\end{align*}

The state vector $|\psi(t)\rangle$ for an arbitrary state describes the time evolution of the whole atom-field system. Here we focus on some analytical properties of the single-mode cavity field, which is obtained from $|\psi(t)\rangle$ by tracing out the atom part
\begin{equation}
	\label{eq10}
	{\rho(t)}_{f} = Tr_{a}[\rho(t)],
\end{equation}
where the subscript $a\,(f)$ denotes the atom (field). We consider  ${\rho(t)}_{f}$  throughout the paper for determining the properties of the field left into the cavity.
%\subsection{Expectation of $\langle a^{\dagger p}a^q\rangle$ w.r.t. $\ket\psi$}
The general form of the expectation $\langle a^{\dagger p}a^q\rangle$ with respect to the state vector $\ket{\psi}$ can be calculated as follows:

%\begin{align}\nonumber
%\label{ex}	
%	& \langle a^{\dagger p}a^q\rangle\\ &=\bra{\psi}a^{\dagger p}a^q\ket{\psi}\\\nonumber
% & = \sum_{m,n}[C_{1,m+1,t}^*\bra{1,m+1} + C_{2,m,t}^*\bra{2,m}]a^{\dagger p}a^q[C_{1,n+1,t}\ket{1,n+1} + C_{2,n,t}\ket{2,n}] \\\nonumber & =
%	\sum_{m,n}\bigg[C_{1,m+1,t}^*\sqrt{\frac{(m+1)!}{(m+1-p)!}}\bra{1,m+1-p} + C_{2,m,t}^*\sqrt{\frac{m!}{(m-p)!}}\bra{2,m-p}\bigg] \\\nonumber& \times \bigg[C_{1,n+1,t}\sqrt{\frac{(n+1)!}{(n+1-q)!}}\ket{1,n+1-q}+ C_{2,n,t}\sqrt{\frac{n!}{(n-q)!}}\ket{2,n-q}\bigg] \\ \nonumber
%	& =\sum_{n}\bigg[C_{1,n+p-q+1,t}^*C_{1,n+1,t}\frac{\sqrt{(n+p-q+1)!(n+1)!}}{(n+1-q)!}+ C_{2,n+p-q,t}^*C_{2,n,t}\frac{\sqrt{(n+p-q)!n!}}{(n-q)!}\bigg]
%\end{align}
%
%
%
%According to the initial condition. If the radiation field is
%in a coherent state then $C_n(0)=e^{-\frac{|\alpha|^2}{2}}\frac{\alpha^n}{\sqrt{n!}}$.
%The state vector $|\psi\rangle$ for an arbitrary state describes the time evolution of
%the whole atom-field system. After we depart the atomic part of $\rho(t)$. We only focus on
%some analytical properties of the single-mode cavity field.
%Now we will consider the field inside the cavity is
%\begin{equation}
%	\label{eq10}
%	{\rho(t)}_{f} = Tr_{a}[\rho(t)],
%\end{equation}
%where the subscript $a\,(f)$ denotes the atom (field). Now we consider  ${\rho(t)}_{f}$  throughout to determine the properties of the field left in the cavity.
%%\subsection{Expectation of $\langle a^{\dagger p}a^q\rangle$ w.r.t. $\ket\psi$}
%We will first find the general form of expectation $\langle a^{\dagger p}a^q\rangle$ for state vector $\ket{\psi}$ as follows:

\begin{eqnarray}\nonumber
\label{ex}	
	\langle a^{\dagger p}a^q\rangle &=& \bra{\psi}a^{\dagger p}a^q\ket{\psi}\\\nonumber
% & = \sum_{m,n}[C_{1,m+1,t}^*\bra{1,m+1} + C_{2,m,t}^*\bra{2,m}]a^{\dagger p}a^q\\\nonumber
%	&\times[C_{1,n+1,t}\ket{1,n+1} + C_{2,n,t}\ket{2,n}] \\\nonumber & =
%	\sum_{m,n}\bigg[C_{1,m+1,t}^*\sqrt{\frac{(m+1)!}{(m+1-p)!}}\bra{1,m+1-p} \\\nonumber
%	&+ C_{2,m,t}^*\sqrt{\frac{m!}{(m-p)!}}\bra{2,m-p}\bigg] \\\nonumber& \times \bigg[C_{1,n+1,t}\sqrt{\frac{(n+1)!}{(n+1-q)!}}\ket{1,n+1-q}\\\nonumber &+ C_{2,n,t}\sqrt{\frac{n!}{(n-q)!}}\ket{2,n-q}\bigg] \\ \nonumber
	& = & \sum_{n}\bigg[C^*_{2,n+p-q+1}(t)C_{2,n+1}(t)\frac{\sqrt{(n+p-q+1)!(n+1)!}}{(n+1-q)!}\\
& & + C^*_{1,n+p-q}(t)C_{1,n}(t)\frac{\sqrt{(n+p-q)!n!}}{(n-q)!}\bigg]
\end{eqnarray}

\section{Nonclassicality criteria under study}

In this section, we derive some criteria to witness the nonclassicality of
the considered quantum state $\ket{\psi}$. While plotting different distribution functions, the numerical values of the atomic transition frequencies and the coupling constant are chosen in such a way that they are in a weak coupling regime $(g \ll \omega_i,\, i=1,2)$, which is the primary assumption for applying RWA. For simplicity, three different forms of $f(\hat{n})$, being an arbitrary polynomial function of $\hat{n}$, are considered. The following plots correspond to the cases $f(n)= \sin({n}),\, 1/\sin({n})$,  and $\ln({n})$. Here, $\sin({n})$ is a bounded and periodic polynomial function, $1/\sin({n})$  is an unbounded, periodic polynomial function and $\ln({n})$ is an unbounded, non-periodic, monotonically increasing polynomial function. In the limiting case $f(\hat{n})=1 $, the Hamiltonian reduces to the basic Jaynes-Cummings model.
\subsection{Photon number distribution}
\begin{figure}[htb]
	\centering
	\includegraphics[scale=1]{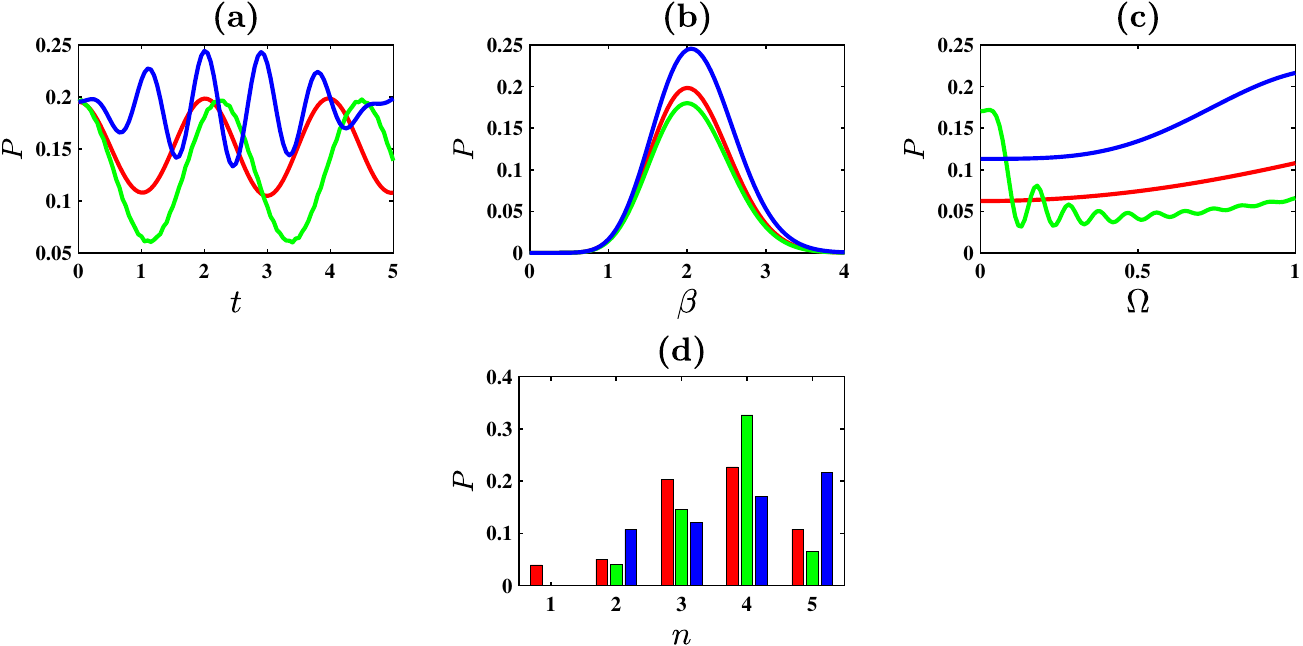}
	\caption{Variation of $p(n)$ with respect to (a) $t$ and for $g=0.5,\,\beta=2,\,\Omega=1 \mbox{GHz},\,\omega_1=\omega_2=100 \mbox{J},\, n=5$, (b) $\beta$ and for $t=1 \mbox{hr},\,g=0.5,\,\Omega=1\mbox{GHz},\,\omega_1=\omega_2=100\mbox{J},\,n=5$, (c) $\Omega$ and for $t=1 \mbox{hr},\, g=0.5,\,\beta=2,\,\omega_1=\omega_2=100\mbox{J},\,n=5 $, (d) $n$ and for $t=1\mbox{hr},\,g=0.5,\,\beta=2,\,\Omega=1\mbox{GHz},\,\omega_1=\omega_2=100\mbox{J}$. Red, green, and blue lines represent $f({n})=\sin({n})$, $1/\sin({n})$, and $\ln({{n}})$ respectively.}
	\label{figpnd}
\end{figure}

Photon number distribution is the probability distribution for finding $n$ photons in a given cavity field. It can be found as the expectation of the density field with respect to the number state $n$ as
\begin{eqnarray}	
	\label{pn}
	p(n) = \langle n |\hat{\rho}|n\rangle
\end{eqnarray}
After tracing out the atom part, the density operator $\hat{\rho}_f$ is given by
\begin{eqnarray}\nonumber
	\label{eq5}
	\hat{\rho}_f & = & |\psi \rangle_f\,{}_f\langle \psi|\\\nonumber
	& = & \sum_{n} \Big[|C_{2,n+1}(t)|^2|n+1\rangle \langle n+1|+ C_{2,n+1}(t) {C^*_{1,n}(t)}|n+1\rangle \langle n|\\
	& &  + {C^*_{2,n+1}(t)} C_{1,n}(t) |n\rangle \langle n+1| +
	|C_{1,n}(t)|^2 |n\rangle \langle n|\Big]
\end{eqnarray}
Substituting \eqref{eq5} into \eqref{pn}, the photon number distribution for the cavity field is obtained as
\begin{eqnarray}\nonumber
	\label{eq6}
	p(n) & = & \langle n|\hat{\rho}|n\rangle\\
	&=&|C_{2,n+1}(t)|^2+|C_{1,n}(t)|^2
\end{eqnarray}
Here \eqref{eq6} gives the probability of finding $n$ photons in a given cavity field state. The simplified form of photon number distribution can be plotted in Fig.~\ref{figpnd} for three different functional values of $f(n)$, namely $\sin(n),\,1/\sin(n)$ and $\ln(n)$.
We can conclude from Fig.~\ref{figpnd} that the variations of $p(n)$ with respect to $t,\,\Omega$ follow wave nature and the amplitude of waves is maximum for $\ln(n)$ and minimum for $1/\sin(n)$. The variation of $p(n)$ with respect to $\beta$ is first increasing and then decreasing and maximum value attained is 0.25. It can be further observed that the peak of $p(n)$ is highest (lowest) for $f(n)=\ln(n)$ $(1/\sin(n))$. While plotting $p(n)$ for different number of photons $n$, all the three functions almost coincide for $n=1$ and have distinct nature for $n>1$.
%
%Further from Fig. \ref{figpnd}, we can conclude that variations of $p(n)$ with respect to $g,\,t,\,\Omega,\,\omega_{1,2}$ follows wave nature along with amplitude of wave nature is largest for $\ln(n)$ and smallest for $1/\sin(n)$ along with medium range of $\sin(n)$. Also the variation of $p(n)$ with respect to $\beta$ is first increasing and then decreasing and maximum value attained is 0.25 and the $p(n)$ is largest(smallest) for, $\sin(n)$ lies in between of two. $p(n)$ for all the three functions almost coincide for $n=1$ and have distinct nature for $n>1$.  value 0.32 approx.

\subsection{Mandel's $Q_M$ parameter}

\begin{figure}[htb]
	\centering
	\includegraphics[scale=1]{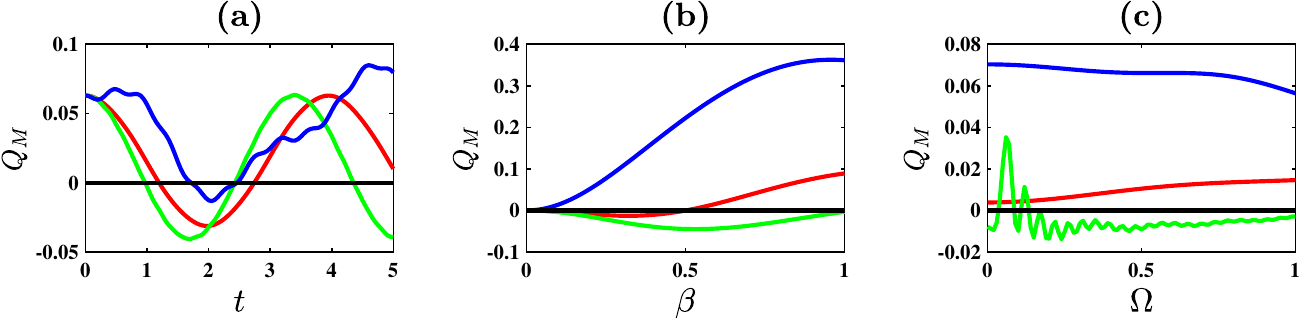}
	\caption{Variation of $Q_M$ with respect to (a) $t$ and for $g=0.5,\,\beta=2,\,\Omega=1\mbox{GHz},\,\omega_1=\omega_2=100\mbox{J}$, (b) $\beta$ and for $t=1\mbox{hr},\,g=0.5,\,\Omega=1\mbox{GHz},\,\omega_1=\omega_2=100\mbox{J}$, (c) $\Omega$ and for $t=1\mbox{hr},\,g=0.5,\,\beta=2,\,\omega_1=\omega_2=100\mbox{J}$. Red, green, and blue lines represent $f({n})=\sin({n})$, $1/\sin({n})$, and $\ln({{n}})$ respectively.}
	\label{figmqf}
\end{figure}

Sub-Poissonian photon statistics are one of the most striking nonclassical characteristics of a quantum system. Mandel's $Q_M$ parameter, which measures how far the photon number distribution deviates from the Poissonian statistics, is used to determine such characteristics \cite{mandel}. $Q_M$ is zero for a coherent field, and the minimal value $Q_M=-1$ is obtained for Fock states, having a well-defined number of photons by definition. The field statistics is sub-Poissonian if $ -1 \leq Q_M < 0$, and thus the phase space distribution cannot be interpreted as a classical probability function. As a result, the negative values of $Q_M$ is sufficient for a field to be nonclassical. But the negativity of $Q_M$ is not a necessary condition to distinguish the quantum states into nonclassical and classical regimes. For example, a state may be nonclassical even though $Q_M$ is positive. If $Q_M>0$, the state's statistics is super-Poissonian but no conclusions can be drawn about their character, for example the $Q_M$ parameter for a well-known nonclassical thermal field is always positive. Next to determine the photon statistics of a single-mode radiation field, we consider the Mandel's $Q_M$
parameter defined by
\begin{equation}
	\label{eq14}
	Q_M=\frac{\langle a^{{\dagger}^2} a^2 \rangle}{\langle a^{\dagger}a\rangle} -\langle a^{\dagger}a\rangle
\end{equation}
Mandel's $Q_M$ function can be plotted for different $f(n)$ such as $\sin{n}$, $1/\sin(n)$, and $\ln(n)$ with respect to different parameters in Fig.~\ref{figmqf}.

Here we can observe that $Q_M$ follows wavy nature with respect to $t$ with maximum amplitude when $f(n)=\ln(n)$. Further $Q_M$ shows nonclassicality for all these three functions with respect to all the parameters along with $\ln(n)$ has minimum nonclassicality whereas $1/\sin(n)$ has maximum nonclassicality. Thus, in general we can conclude that the considered quantum state is nonclassical in nature for different polynomial functional values of $f(n)$ and order of nonclassicality may increase or decrease according to the choice of $f(n)$. Also the nonclassicality decreases with increasing of $\beta$, $\Omega$ and wavy in nature while increasing $t$.

\subsection{Wigner function}
The nonclassicality of a quantum state can be studied in terms of its phase-space distribution characterized by the Wigner function. For a quantum state $\hat{\rho}$, the Wigner function of the system is defined in terms of the coherent state basis as \cite{ql,pathak,chatterjee}
\begin{eqnarray*}
	W(\gamma, \gamma^*) = \frac{2}{\pi^2}e^{2|\gamma|^2} \int d^2\delta{\langle-\delta|\hat{\rho}|\delta\rangle e^{-2(\beta^*\delta-\beta\delta^*)}},
\end{eqnarray*}
where $|\delta\rangle=\exp(-|\delta|^2/2+\delta \hat{a}^\dag)|0\rangle$ is a coherent state. By using the relation \cite{abramowitz72}
\begin{eqnarray*}
	\sum_{n=k}^\infty n_{C_k}\,y^{n-k} = (1-y)^{-k-1},
\end{eqnarray*}
the Wigner function can be expressed in series form as \cite{moyacessa93}
\begin{eqnarray}
	\label{eqn8}
	W(\gamma, \gamma^*) = \frac{2}{\pi} \sum_{k=0}^\infty (-1)^k \langle \gamma,k|\hat{\rho}|\gamma,k\rangle ,
\end{eqnarray}
where $|\gamma,k\rangle$ is the usual displaced number state. The partial negative value of the Wigner function is a one-sided condition for the nonclassicality of the related state \cite{wangz},
in the sense that one cannot conclude the state is classical
when the Wigner function is positive everywhere. For example,
the Wigner function of the squeezed state is Gaussian and
positive everywhere but it is a well-known nonclassical state.
For a classical state, a necessary but not sufficient condition
is the positivity of the Wigner function. Hence, a state with a
negative region in the phase-space quadrature is essentially
nonclassical.
Now the displaced number state $|\gamma, k\rangle$ can be expressed in a number state basis as
\begin{eqnarray}
	\label{eqn9}
	|\gamma, k\rangle & = &
	D(\gamma)| k \rangle\\\nonumber
	& = & e^{-\frac{|\gamma|^2}{2}}\sum_{m=0}^{k} \frac{ \gamma^{*{m}}}{m!} \sqrt{\frac{k!}{(k-m)!}}\sum_{p=0}^{\infty} \frac{ \gamma^p}{p!}\sqrt{\frac{(k-m+p)!}{(k-m)!}}|k-m+p \rangle
\end{eqnarray}
Thus
\begin{eqnarray}\nonumber
\label{eqne}
	& & \langle \gamma,k|n \rangle\\\nonumber & = &
	e^{-\frac{|\gamma|^2}{2}}\sum_{m=0}^{k} \frac{(-\gamma)^{m}}{m!} \sqrt{\frac{k!}{(k-m)!}}
\sum_{p=0}^{\infty}\frac{ \gamma^{*{p}}}{p!}  \sqrt{\frac{(k-m+p)!}{(k-m)!}}\langle k-m+p|n\rangle\\
	%& = & e^{-\frac{|\gamma|^2}{2}}\sum_{m=0}^{k} \frac{(-\gamma)^{m}}{m!} \sqrt{\frac{k!}{n!}} \frac{(\gamma^*)^{n-k+m} n!}{(k-m)!(n-k+m)!}\\\nonumber
	%& = & e^{-\frac{|\gamma|^2}{2}} \sqrt{\frac{k!}{n!}} (\gamma^*)^{n-k}\sum_{m=0}^{k}  \frac{(-|\gamma|^2)^{m} n!}{(k-m)!m!(n-k+m)!}\\
	& = & e^{-\frac{|\gamma|^2}{2}} \sqrt{\frac{k!}{n!}} (\gamma^*)^{n-k}L_{k}^{(n-k)}\left(|\gamma|^2 \right)
\end{eqnarray}
where $L_{m}^{(k)}(x)=\sum_{n=0}^{m}  \frac{(-x)^{n} (m+k)!}{(m-n)! n!(k+n)!}$ is the associated Laguerre polynomial \cite{agg}. Substituting \eqref{eqne} into \eqref{eqn8} we get
\begin{eqnarray}\nonumber
	 W(\gamma, \gamma^*)
%& = & \frac{2}{\pi} \sum_{k=0}^{\infty} (-1)^k \sum_{n=0}^{\infty}\bigg[|c_{a,n}|^2 \langle\gamma,k|n\rangle\langle n|\gamma,k \rangle\\\nonumber
%	& &  + c_{a,n}c_{b,n+1}^* \langle \gamma,k|n \rangle \langle n+1|\gamma,k \rangle\\\nonumber
%	& & + c_{a,n}^*c_{b,n+1} \langle\gamma,k|n+1\rangle \langle n|\gamma,k\rangle\\\nonumber
%	& & +|c_{b,n+1}|^2 \langle \gamma,k|n+1 \rangle \langle n+1|\gamma,k\rangle\bigg]\\\nonumber
	& = & \frac{2e^{-|\gamma|^2}}{\pi} \sum_{k=0}^{\infty} (-1)^k k!\sum_{n=0}^{\infty} \bigg[|c_{a,n}|^2 \frac{|\gamma|^{2(n-k)}}{n!} \left\{L_{k}^{(n-k)} \left(|\gamma|^2 \right) \right\}^2\\
	& & +|c_{b,n+1}|^2  \frac{|\gamma|^{2(n+1-k)}}{(n+1)!} \left\{L_{k}^{(n+1-k)} \left(|\gamma|^2 \right) \right\}^2\bigg]
\end{eqnarray}
\begin{figure}[htb]
	\centering
	\includegraphics[scale=.7]{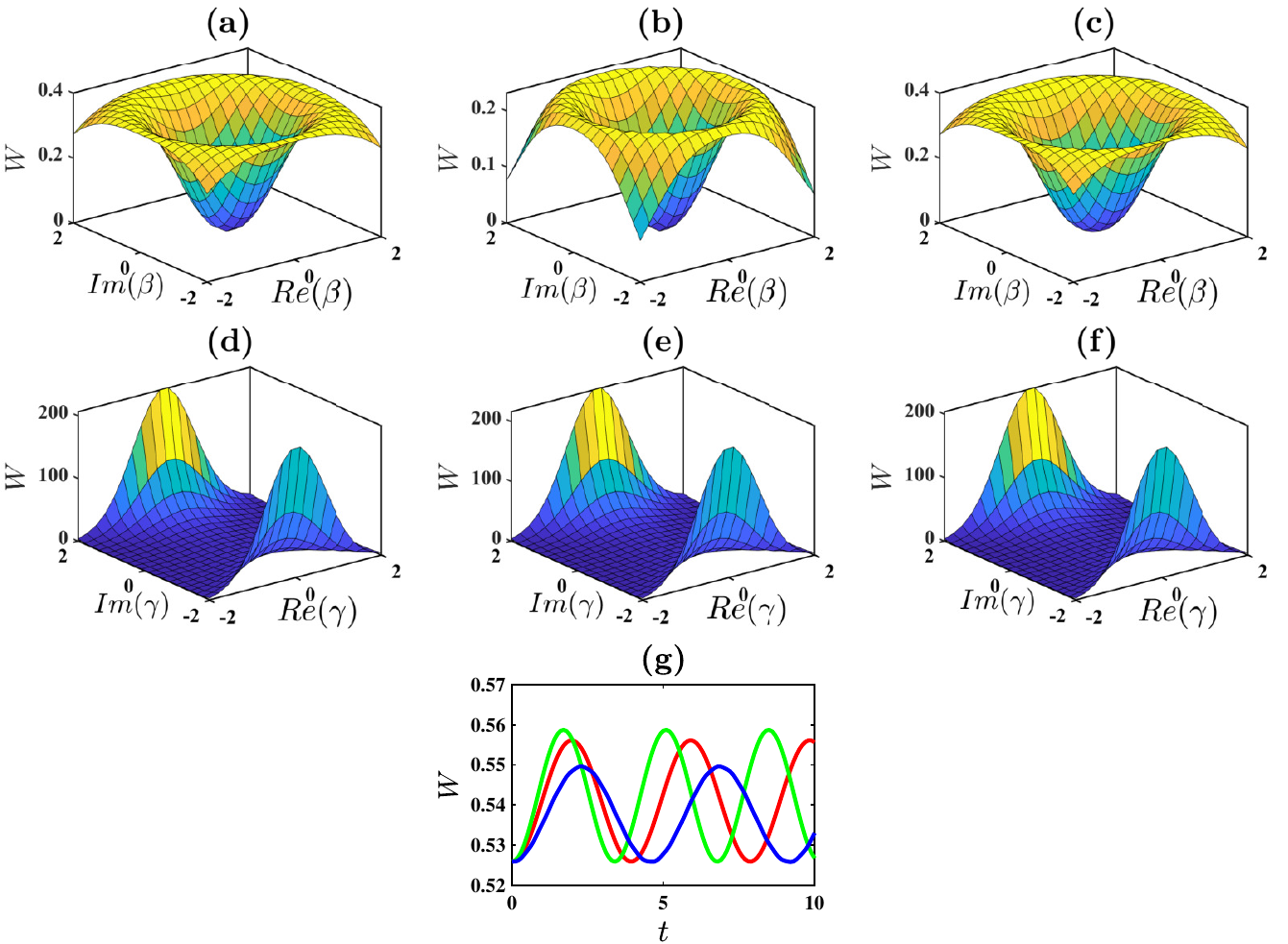}
	\caption{Variation of $W$ with respect to $\beta$ ($\gamma$) in the first (second) row and for $f(n)=\sin(n)$ in (a) and (d), $1/\sin{n}$ in (b) and (e), $\ln{n}$ in (c) and (f). $W$ is plotted with respect to $t$ in (g). Other parameters are the same as in Fig.~\ref{figpnd}.}
	\label{figwf}
\end{figure}
Here it is clear that the Wigner function shows the non-gaussian \cite{priyalp} behaviour with respect to real and imaginary $\beta$ and $\gamma$. As the Wigner function has no negative region, we can conclude that this function is not showing nonclassicality. Also the variation of Wigner function with respect to the parameter $t$ shows wavy nature.

\subsection{Anti-bunching}
\begin{figure}[htb]
	\centering
	\includegraphics[scale=1]{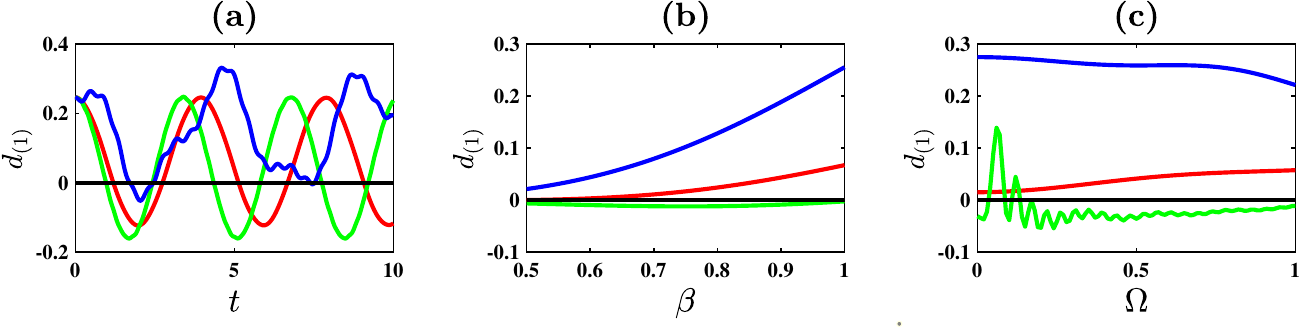}
	\caption{Variation of $d_{(1)}$ for different $f(n)$ and with respect to (a) $t$, (b) $\beta$ and (c) $\Omega$. Other parameters are the same as in Fig.~\ref{figpnd}.}
	\label{figlab}
\end{figure}

  The theory of majorization in \cite{lee, path} gives us the expression of anti-bunching in a quantum state which satisfies the condition for nonclassicality as
\begin{equation*}
	\label{eq14}
	d_{(1)}=\langle a^{{\dagger}^2} a^2 \rangle -{\langle a^{\dagger}a\rangle}^2
	<0
\end{equation*}
The negativity of $d_{(1)}$ implies that the probability of detecting independent photons is more than the clustered photons. We can find the expression of $d_{(1)}$ by using the expectation values of $a^{{\dagger}^2} a^2$ and $a^{\dagger}a$ from \eqref{ex}. Also the anti-bunching criteria can be plotted in Fig.~\ref{figlab} with respect to different parameters. It is seen that anti-bunching has the similar behaviour as $Q_M$.

\subsection{Squeezing properties}
In order to analyze the quantum fluctuations of the field quadratures, we consider two Hermitian operators which are combinations of photon creation and annihilation operators as
$${x}=\frac{{a}+{a}^{\dagger}}{2},\,\,\,\,\,\,{p}=\frac{{a}-{a}^{\dagger}}{2i}$$\\
with the commutation relation $\left[{x},{p}\right] = i/2.$ They obey the Heisenberg uncertainty relation of the form
$\langle(\Delta{x})^2\rangle\langle(\Delta{p})^2\rangle\geq\frac{1}{16}$, and thus the quadrature squeezing occurs whenever $\langle(\Delta{x})^2\rangle<\frac{1}{4}$ or $\langle(\Delta{p})^2\rangle<\frac{1}{4}$. It is convenient to introduce the squeezing parameters as \cite{a10}
\begin{eqnarray}\nonumber
	s_x & = & 4\langle(\Delta{x})^2\rangle-1\\\nonumber
	& = & 2\langle{a}^\dag{a}\rangle+\langle{a}^2\rangle+\langle{a}^{\dag2}\rangle-
	\langle{a}\rangle^2-\langle{a}^\dag\rangle^2-2\langle{a}\rangle\langle{a}^\dag\rangle,
\end{eqnarray}
and
\begin{eqnarray}\nonumber
	s_p & = & 4\langle(\Delta{p})^2\rangle-1\\\nonumber
	& = & 2\langle{a}^\dag{a}\rangle-\langle{a}^2\rangle-\langle{a}^{\dag2}\rangle+
	\langle{a}\rangle^2+\langle{a}^\dag\rangle^2-2\langle{a}\rangle\langle{a}^\dag\rangle,
\end{eqnarray}
such that the squeezing exists in ${x}$ or ${p}$ quadrature if $-1<s_x<0$ or $-1<s_p<0$, respectively. The negativity of squeezing is not a necessary condition but a sufficient one. The expectations can be calculated using \eqref{ex}.
\begin{figure}[htb]
	\centering
	\includegraphics[scale=1]{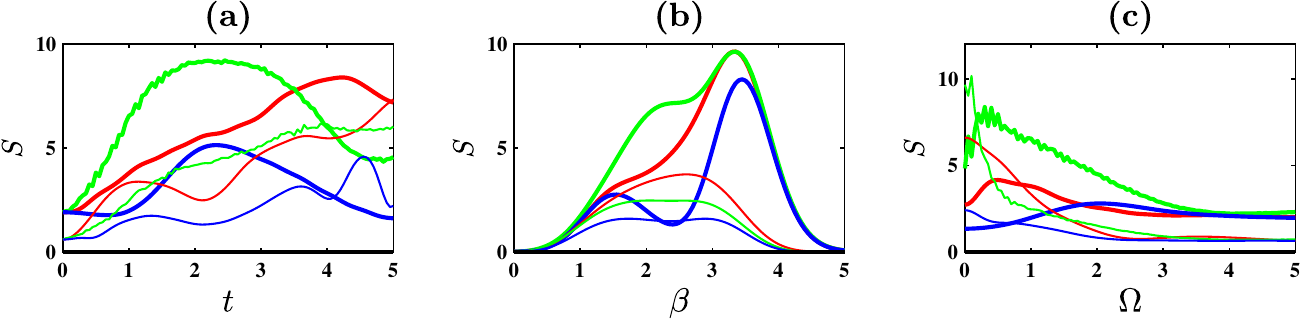}
	\caption{Variation of Squeezing parameters $S=s_{x}$ (thick line) and $s_{p}$ (thin line) with respect to $t$ in (a), $\beta$ in (b) and $\Omega$ in (c). Other parameters are the same as in Fig.~\ref{figpnd}.}
	\label{figls}
\end{figure}
In Fig.~\ref{figls} we observe that the squeezing parameters $S=s_x$, $s_p$ are positive for all parametric values of $t$, $\beta$, and $\Omega$ and thus they fail to detect the nonclassical nature of the cavity field.

\subsection{$Q$ Function}
The uncertainty principle prevents a direct phase-space description of a quantum mechanical system. This fact leads to the creation of quasiprobability distributions, which are very useful in quantum mechanics because they provide a quantum-classical correspondence and make it easier to calculate quantum mechanical averages in a way that is similar to the calculation of classical phase-space averages \cite{q1}. The $Q$ function is one such quasiprobability distribution, and its zeros are a sign of nonclassicality \cite{q2}. The $Q$ function is calculated as follows:
\begin{align*}
Q&=\frac{1}{\pi}\bra{\alpha}{\rho}\ket{\alpha}\\
&=\bra{\alpha}
	\sum_{n,p}[C_{1,n+1} C^*_{1,p+1}|n+1\rangle  \bra{p+1} + C_{2,n}C^*_{2,p} |n \rangle  \bra{p}]\ket{\alpha}\\
	&=
	\sum_{n,p}[C_{1,n+1} C^*_{1,p+1}e^{\frac{-|\alpha|^2}{2}}\frac{\alpha^{*(n+1)}}{\sqrt{(n+1)!}}e^{\frac{-|\alpha|^2}{2}}\frac{\alpha^{p+1}}{\sqrt{(p+1)!}}+C_{2,n}C^*_{2,p}e^{\frac{-|\alpha|^2}{2}}\frac{\alpha^{*(n)}}{\sqrt{(n)!}}e^{\frac{-|\alpha|^2}{2}}\frac{\alpha^{p}}{\sqrt{p!}} \\
	&=
	e^{-|\alpha|^2}\sum_{n,p}\left[\frac{\alpha^{*n}\alpha^{p}}{\sqrt{p!n!}}\left\lbrace C_{1,n+1} C^*_{1,p+1}\frac{|\alpha|^2}{\sqrt{(p+1)(n+1)}}+C_{2,n}C^*_{2,p}\right\rbrace\right]
\end{align*}
%\begin{align*}
%\bra{\beta}\ket{n}&=e^{\frac{-|\beta|^2}{2}}\frac{\beta^{*n}}{n!}\\
%\end{align*}
\begin{figure}[htb]
	\centering
	\includegraphics[scale=.8]{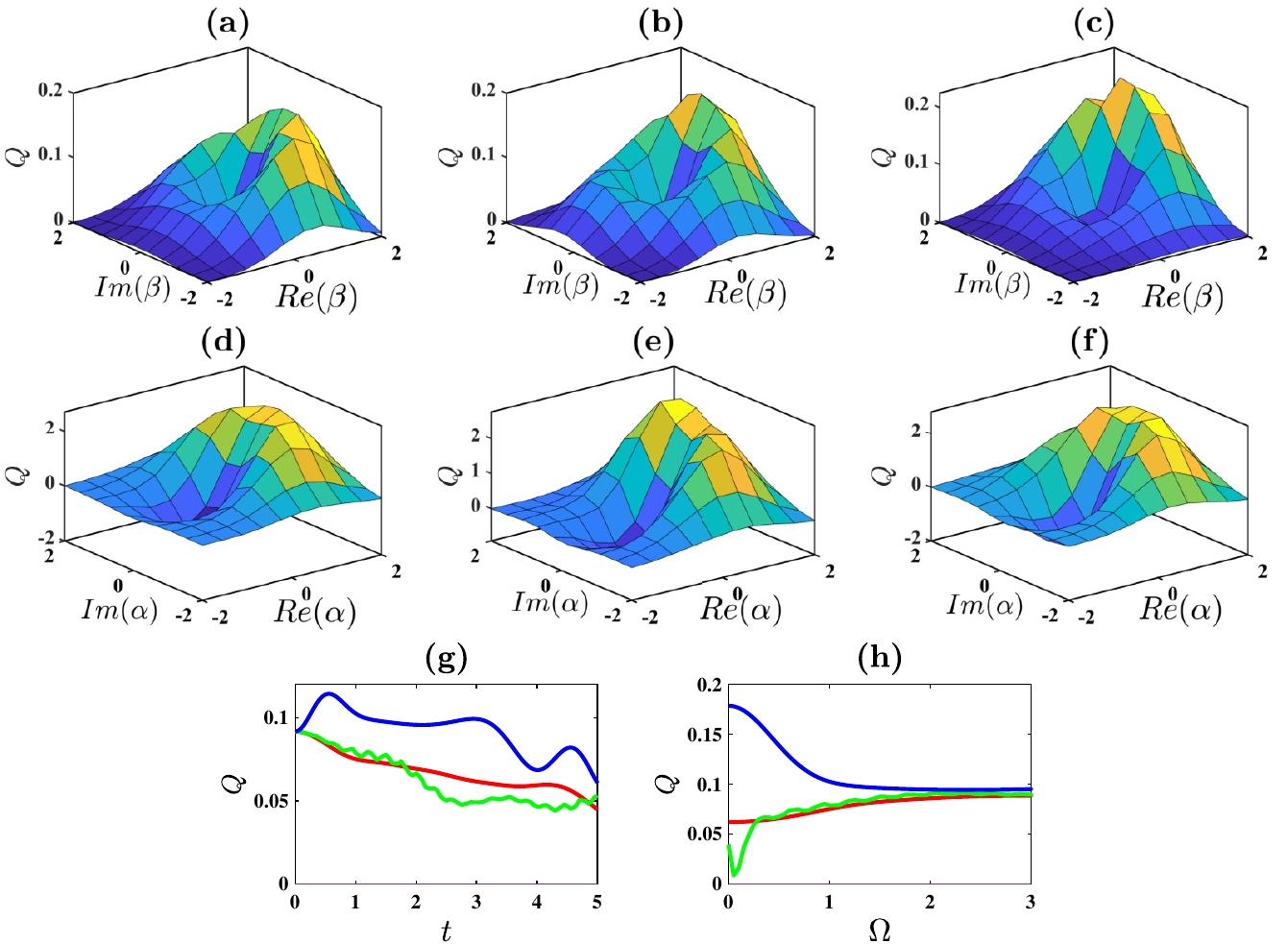}
	\caption{Variation of $Q$ with respect to $\beta$ ($\alpha$) in the first (second) row and for $f(n)=\sin(n)$ in (a) and (d), $1/\sin{n}$ in (b) and (e), $\ln{n}$ in (c) and (f). $Q$ is plotted with respect to $t$ in (g) and $\Omega$ in (h). Other parameters are the same as in Fig.~\ref{figpnd}.}
	\label{figqf}
\end{figure}
In Fig.~\ref{figqf}, we observe that the $Q$ distribution shows the wavy and non-Gaussian nature and remains positive everywhere.

\section{Conclusion}
In this article, we have considered a deformed model of interaction between a two-level atom and a single-mode electromagnetic cavity field. Assuming that the atom enters the cavity in its excited state $\ket{2}$ and the field is initially in a coherent state, we have derived the cavity field state. For this cavity field state, we have found some statistical properties like photon number distribution, Wigner function, Mandel's $Q_M$ parameter, anti-bunching, squeezing parameter, and $Q$-function. All of these criteria have specific conditions for witnessing nonclassicality such as the negativity of the lower-order antibunching and the Poissonian statistics of Mandel’s $Q_M$ parameter are used to reveal the nonclassicality of the quantum field state. Here we have observed that the $Q$-function, Wigner function, and squeezing parameter failed to show nonclassicality.

\begin{center}
	\textbf{ACKNOWLEDGEMENT}

\end{center}
Naveen and Deepak acknowledge the financial support from the Council of Scientific and Industrial Research, Govt. of India (Grant no.
09/1256(0004)/2019-EMR-I and 09/1256(0006)/2019-EMR-I, respectively).


\begin{thebibliography}{99}

\bibitem{jcm} E. T. Jaynes, F. W. Cummings, Proc. IEEE. 51, 89 (1963).
\bibitem{rad} B. W. Shore, P. L. Knight, J. Mod. Opt. 40, 1195 (1993).
\bibitem{it} D. Bouwmeester, A. Ekert, A. Zeilinger, \textit{The Physics of Quantum Information}, Berlin: Springer (2000).
\bibitem{d} E. Knill, R. Laflamme, G. J. Milburn, Nature, 409, 46 (2001).
\bibitem {brunal} C. Brunel, B. Lounis, P. Tamarat, M. Orrit, Phys. Rev. Lett. 83, 2722 (1999).
\bibitem{hagley} X. Maître, E. Hagley, G. Nogues, C. Wunderlich, P. Goy, M. Brune, J.M. Raimond, S. Haroche, Phys. Rev. Lett. 79, 769 (1997).
\bibitem {henrich} M. Hennrich, T. Legero, A. Kuhn, G. Rempe, Phys. Rev. Lett. 85, 4872 (2000).
\bibitem{ql} M. O. Scully, M. S. Zubairy, \textit{Quantum Optics}, Cambridge University Press (1997).
%\bibitem{a1d} Dhar H.S.; Banerjee S.; Chatterjee A.; Ghosh R. Ann. Phys. 2013, 331, 97.
\bibitem{qi} S. L. Braunstein,  P. van Loock, Rev. Mod. Phys. 77, 513–577 (2005).
\bibitem {pn} W. Schleich, J. A. Wheeler, Nature 326, 574 (1987).
\bibitem{siva} S. Sivakumar, Int. J. Theor. Phys. 43, 2405-2421 (2004).
\bibitem{deformed} C. Cohen-Tannoudji,  J. Dupont-Roc, G. Grynberg, \textit{Atom-Photon Interactions; Basic Processes and Applications}, John
Wiley and Sons, New York (1992).
\bibitem{deformed1} V. I. Mańko, G. Marmo, E. C. G. Sudarshan, F. Zaccaria, Phys. Scr. 55, 528 (1997).
\bibitem{deformed2} R. A. Zait, Phys. Lett. 319, 461 (2003).
\bibitem{filo} R. L. de Matos Filo, W. Vogel, Phys. Rev. A 58, 2326 (1998).
\bibitem{mandel} L. Mandel, Opt. Lett. 4, 205 (1979).
\bibitem{pathak} P. K. Pathak, G. S. Agarwal, Phys. Rev. A 71, 043823 (2005).
\bibitem{chatterjee} A. Chatterjee, J. Mod. Opt. 59 (9), 814 (2012).
\bibitem{abramowitz72} M. Abramowitz, I. A. Stegun, \textit{Handbook of Mathematical Functions}, New York: Dover (1972).
\bibitem{moyacessa93} H. Moya-Cessa P. L. Knight, Phys. Rev. A 48, 2479 (1993).
\bibitem{wangz} Z. Wang, H. C. Yuan, H. Y. Fan, J. Opt. Soc. Am. B 28, 1964 (2011).
\bibitem{agg} G. S. Agarwal, K. Tara, Phys. Rev. A 43(1), 492 (1991).
\bibitem{priyalp} P. Malpani, K.thapliyal, J. Banerji, A. Pathak, arxiv:2109.12145v1 (2021).
\bibitem{lee} C. T. Lee, Phys. Rev. A 41, 1721 (1990).
\bibitem{path} A. Pathak, M. E. Garcia, Appl. Phys. B 84, 479 (2006).
\bibitem{a10} A. Chatterjee, R. Ghosh, J. Opt. Soc. Am. B 33(7), 1511 (2016).
\bibitem{q1} K. Thapliyal, S. Banerjee, A. Pathak, S. Omkar, V. Ravishankar, Ann. Phys. 362, 261 (2015).
\bibitem{q2} N. Lütkenhaus, S. M. Barnett, Phys. Rev. A 51, 3340 (1995).




%\bibitem{jcm}  Jaynes E.T.; Cummings F.W. Proc. IEEE 1973, 51, 89.
%\bibitem{rad} Shore B.W.; Knight P.L. J. Mod. Opt. 1993, 40, 1195.
%\bibitem{it} D. Bouwmeester, A. Ekert A and A. Zeilinger, \textit{The Physics of Quantum Information (Berlin: Springer)} (2000).
%\bibitem{d} E. Knill, R. Laflamme, G.J. Milburn, Nature (London) 409 (2001) 46.
%\bibitem {brunal}C. Brunel, B. Lounis, P. Tamarat, M. Orrit, Phys. Rev. Lett. 83 (1999) 2722.
%\bibitem{hagley} X. Maître, E. Hagley, G. Nogues, C. Wunderlich, P. Goy, M. Brune, J.M. Raimond, S. Haroche, Phys. Rev. Lett. 79 (1997) 769.
%\bibitem {henrich}M. Hennrich, T. Legero, A. Kuhn, G. Rempe, Phys. Rev. Lett. 85 (2000) 4872.
%%\bibitem{a1d} Dhar H.S.; Banerjee S.; Chatterjee A.; Ghosh R. Ann. Phys. 2013, 331, 97.
%\bibitem{ql}Scully, M.O.; Zubairy, M.S. Quantum Optics;
%Cambridge University Press: UK, 1997.
%\bibitem{qi}Braunstein, S.L.; van Loock, P. Rev. Mod. Phys. 2005,
%77, 513–577
%%\bibitem{scully97} M. O. Scully and M. S. Zubairy, \textit{Quantum Optics, Cambridge University Press}, (1997).
%
%\bibitem {pn} W. Schleich and J. A. Wheeler, Nature \textbf{326}, 574 (1987);
%J. Opt. Soc. Am. B \textbf{4}, 1715 (1987).
%\bibitem{siva} S. Sivakumar, Int. J. Theor. Phys. \textbf{43} 2405-2421 (2004).
%\bibitem{deformed} Cohen-Tannoudji C.; Dupont-Roc J.; Grynberg G. Atom-Photon Interactions: Basic Processes and Applications; John
%Wiley and Sons: New York, 1992.
%\bibitem{deformed1} Mańko V.I.; Marmo G.; Sudarshan E.C.G.; Zaccaria F.
%Phys. Scr. 1997, 55, 528.
%\bibitem{deformed2}Zait R.A. Phys. Lett. 2003, 319, 461.
%%\bibitem{am12} A Moroz, Europhys. Lett. 100(2012) 60010.
%%\bibitem{am14}A Moroz, Ann. Phys. 340(2014) 252-266.
%\bibitem{mandel} L. Mandel, Opt. Lett., Vol. 4, 205 (1979).
%\bibitem{pathak} P. K. Pathak, and G. S. Agarwal, Phys. Rev. A, Vol. 71, 043823 (2005).
%\bibitem{chatterjee} A. Chatterjee, J. Mod. Opt., Vol. 59 (9), 814 (2012)
%\bibitem{abramowitz72} M. Abramowitz and I. A. Stegun, \textit{Handbook of Mathematical Functions}, New York: Dover (1972).
%\bibitem{moyacessa93} H. Moya-Cessa, and P. L. Knight, Phys. Rev. A, Vol. 48, 2479 (1993).
%\bibitem{wangz} Z. Wang, H. C. Yuan, and H. Y. Fan, J. Opt. Soc. Am. B, Vol. 28, 1964 (2011).
%\bibitem{agg} G. S. Agarwal, and K. Tara, Phys. Rev. A, Vol. 43 (1), 492 (1991).
%\bibitem{priyalp}P. Malpani, K.thapliyal, J. Banerji, A. Pathak, arxiv:2019.12145v1 (2021).
%\bibitem{lee} C. T. Lee, Phys. Rev. A 41, 1721 (1990).
%\bibitem{path} A. Pathak and M. E. Garcia, Appl. Phys. B 84, 479–484 (2006).
%\bibitem{a10} A. Chatterjee, and R. Ghosh, J. Opt. Soc. Am. B, Vol.
%33(7), 1511 (2016).
%\bibitem{q1} K. Thapliyal, S. Banerjee, A. Pathak, S. Omkar, and V. Ravishankar, Ann. Phys. 362, 261–286 (2015).
%\bibitem{q2} N. Lütkenhaus and S. M. Barnett, Phys. Rev. A 51, 3340 (1995).
%\bibitem{machida} S.R. Friberg, S. Machida, Y. Yamamoto, Phys. Rev. Lett. 69 3165 (1992).
%\bibitem{zeilinger98} A. Zeilinger, Rev. Mod. Phys. \textbf{71}, S288 (1998).
%\bibitem{braunstein05} S. L. Braunstein and P. van Loock, Rev. Mod. Phys. \textbf{77}, 513 (2005).
%\bibitem{naik00} D. S. Naik, C. G. Peterson, A. G. White, A. J. Berglund and P. G. Kwiat, Phys. Rev. Lett. \textbf{84}, 4733 (2000).
%\bibitem{bouwmeester98} D. Bouwmeester, J.-W. Pan, M. Daniell, H. Weinfurter, M. Zukowski and A. Zeilinger, Nature \textbf{394}, 841 (1998).
%\bibitem{brune96} M. Brune, E. Hagley, J. Dreyer, X. Maître, A. Maali, C. Wunderlich, J. M. Raimond and S. Haroche, Phys. Rev. Lett. \textbf{77}, 4887 (1996).
%\bibitem{zavatta04} A. Zavatta, S. Viciani, and M. Bellini, Science \textbf{306}, 660 (2004).
%\bibitem{ourjoumtsev07} A. Ourjoumtsev, A. Dantan, R. Tualle-Brouri and Ph. Grangier, Phys. Rev. Lett. \textbf{98}, 030502 (2007).
%\bibitem{browne03} D. E. Browne, J. Eisert, S. Scheel, and M. B. Plenio, Phys. Rev. A \textbf{67}, 062320 (2003).
%\bibitem{garcia04} R. Garc\'{i}a-Patr\'{o}n, J. Fiur\'{a}\v{s}ek, N. J. Cerf, J. Wenger, R. Tulle-Brouri and P. Grangier, Phys. Rev. Lett. \textbf{93}, 130409 (2004).
%\bibitem{nha04} H. Nha and H. J. Carmichael, Phys. Rev. Lett. \textbf{93}, 020401 (2004).
%\bibitem{bartlett02} S.D. Bartlett and B. C. Sanders, Phys. Rev.A \textbf{65}, 042304 (2002).
%\bibitem{knill01} E. Knill, R. Laflamme and G. J. Milburn, Nature (London) \textbf{409}, 46 (2001).
%\bibitem{brunel99} C. Brunel, B. Lounis, P. Tamarat and Michel Orrit, Phys. Rev. Lett. \textbf{83}, 2722 (1999).
%\bibitem{friberg92} S. R. Friberg, S. Machida, and Y. Yamamoto, Phys. Rev. Lett. \textbf{69}, 3165 (1992).
%\bibitem{maitre97} X. Ma\^{\i}tre, E. Hagley, G. Nogues, C. Wunderlich, P. Goy, M. Brune, J. M. Raimond and S. Haroche, Phys. Rev. Lett. \textbf{79}, 769  (1997).
%\bibitem{henrich00} M. Hennrich, T. Legero, A. Kuhn and G. Rempe, Phys. Rev. Lett. \textbf{ 85}, 4872 (2000).
%\bibitem{alsing} P. Alsing, D. S. Guo, and H. J. Carmichael, Phys. Rev. A 45, 5135 (1992).
%\bibitem{biao} S. B. Zheng, Phys. Rev. A 74, 043803 (2006).
%\bibitem{arpita} A. Ghosh, and P. K. Das, Int. J. Theor. Phys. 47, 1731 (2008).
%\bibitem{chen} C. Y. Chen, M. Feng, and K. L. Gao, Phys. Rev. A  73, 034305 (2006).
%\bibitem{peng1} J. S. Peng, G. X. Li, and P. Zhou, Phys. Rev. A, Vol. 46 (3), 1516 (1992).
\end{thebibliography}
\end{document}